# Anchoring-Based Causal Design (ABCD): Estimating the Effects of Beliefs


Raanan Sulitzeanu-Kenan, Micha Mandel and Yosef Rinott

The Hebrew University of Jerusalem


August 2025


A central challenge in any study of the effects of beliefs on outcomes, such as decisions and behavior, is the risk of omitted variables bias. Omitted variables—frequently unmeasured or even unknown—can induce correlations between beliefs and decisions that are not genuinely causal, in which case the omitted variables are referred to as confounders. To address the challenge of causal inference, researchers frequently rely on information provision experiments to randomly manipulate beliefs. The information supplied in these experiments can serve as an instrumental variable (IV), enabling causal inference, so long as it influences decisions exclusively through its impact on beliefs. However, providing varying information to participants to shape their beliefs can raise both methodological and ethical concerns. Methodological concerns arise from potential violations of the exclusion restriction assumption. Such violations may stem from information source effects – when attitudes toward the source affect the outcome decision directly, thereby introducing a confounder. An ethical concern arises from manipulating the provided information, as it may involve deceiving participants. This paper proposes and empirically demonstrates a new method for treating beliefs and estimating their effects – the Anchoring-Based Causal Design (ABCD) – which avoids deception and source influences. ABCD combines the cognitive mechanism known as anchoring with instrumental variable (IV) estimation. Instead of providing substantive information, the method employs a deliberately non-informative procedure in which participants compare their self-assessment of a concept to a randomly assigned anchor value. We present the method and the results of eight experiments demonstrating its application, strengths, and limitations. We conclude by discussing the potential of this design for advancing experimental social science.



Acknowledgements: We thank Donald Green, Moses Shayo, Matthew Simonson, David Weisburd, Chagai Weiss, Christopher Wlezien, Omer Yair and participants of seminars at the Hebrew University, and the 1st MethodsNet Conference (Louvain de-Nueve, 2024) for valuable comments. The Center for Interdisciplinary Data Science Research (CIDR) at the Hebrew University provided generous financial support.




Standard economic theory and evolutionary biological models describe human decisions as a combination of three factors: preferences, constraints, and beliefs (Gintis, 2006; Haaland et al., 2023). This paper introduces a new experimental method for identifying the causal effect of beliefs on decisions, by selectively treating beliefs using anchoring, holding other factors, whether observed or unobserved, constant.

A central challenge in causal analysis is the presence of confounders. In the case of the effect of a certain belief on a particular decision, confounders are variables, such as personal characteristics or ecological conditions, that jointly influence the belief and the decision in question. For instance, in examining whether beliefs about a descriptive social norm of charitable donation (measured by people's beliefs about the mean yearly household donation to charity) affect donation decision, factors such as an individual's financial status or personality traits, like prosocial tendencies, thus influence both correlations between beliefs and decisions that are not causal.

Experimental methods address the general problem of confounders by randomly treating the factor of interest. In social sciences, priming experiments regularly serve as a method for treating preferences (Cohn & Maréchal, 2016), and information provision experiments serve the purpose of treating beliefs and (perceived) constraints. A researcher who wishes to test the effect of beliefs regarding the social norm of charitable donation on donation decisions, would therefore randomly provide varying information (e.g., expert reports) regarding social donation norms.

However, altering the information of experimental participants involves two major challenges. First, information is often associated, either explicitly or implicitly, with a source (e.g., the expert or institution that generated the information). Participants



may generate attitudes toward the source, which may affect the outcome decision directly, thereby introducing a new confounder—referred to as information source effect (see Haaland et al., 2023; Stantcheva, 2023). Another challenge of presenting participants with differing information may raise ethical concerns of deception, particularly if the information is perceived as misleading or manipulative.

Addressing these challenges requires considerable endeavors. Researchers may randomize the source of the information across several potential sources to mitigate information source effects. Yet this procedure requires to identify a set of sufficiently credible sources that properly balances the potential confounding effect. To avoid deception researchers may either use true information in one experimental arm and no information in the other, or use varying true estimates when addressing beliefs regarding factually undetermined information (e.g., the likelihood of a recession, or the expected competitiveness of an election). Using 'no information' as reference may violate the exclusion restriction as the difference in treatment is not restricted to the content of the information but also to its availability. Providing varying yet true informative estimates is challenging as obtaining such estimates is often difficult, and their variability is typically limited.

To address these challenges, we propose and demonstrates the efficacy of a novel experimental method for treating beliefs and estimating their effects – the Anchoring-Based Causal Design (ABCD). This approach eliminates source-related confounding and completely avoids the ethical concerns associated with deception. ABCD leverages the cognitive mechanism known as *anchoring*: the tendency for individuals to adjust their estimates of a quantitative value based on an initially presented reference point (Tversky



& Kahneman, 1974; for a review see: Furnham & Boo, 2011). To illustrate, consider once again a study of the effect of beliefs regarding the social norm of charitable donation. In an ABCD design, a researcher merely suggests a number (*a*) by asking respondents, "Do you think the mean yearly household donation to charity is higher or lower than *a*?". By suggesting *a* as a possible value of the belief, the researcher can effectively treat the respondent's belief upward or downward towards *a* without providing new information, and without any deception.

**The anchoring-based causal design (ABCD)**

ABCD is a method for treating beliefs by using anchoring as an instrumental variable (IV) for causal inference, with the goal of avoiding both confounding variable bias and the ethical concerns that arise in information provision experiments. We begin with a brief review of the anchoring mechanism, followed by a detailed description of the anchoring-based treatment, and its integration within instrumental variable estimation.

*Anchoring*

When people make estimates about a particular concept they often start from an initial value, which may be externally suggested, and adjust from this value to determine the final estimate. However, adjustments are typically insufficient, and the initially considered value (the anchor) influences the final estimate, an effect known as anchoring. This phenomenon was first identified by Sherif, Taub, and Hovland (1958). Tversky and Kahneman (1974) demonstrated anchoring in a classic study, in which participants were



requested to consider whether the percentage of African nations within the United Nations is larger or smaller than a result of a wheel of fortune rigged to produce only the values 10 or 65 (the comparative judgment stage). Next, they were asked to estimate the numerical answer (the absolute judgment stage). Participants who were exposed to the higher (lower) value in the comparative judgment stage, provided a higher (lower) estimate for the percentage of African nations in the UN in the absolute judgment stage. Since then, numerous studies have demonstrated the robustness and sizable qualities of anchoring across diverse settings and contexts (Furnham & Boo, 2011; Schley, 2023).

Several theories have been proposed to explain this phenomenon. The Insufficient Adjustment Model posits that people begin with the anchor and make incremental adjustments that typically fall short of reaching an anchorless estimate (Epley & Gilovich, 2005; Tversky & Kahneman, 1974). The Numeric Priming Model suggests that anchors enhance the accessibility of nearby values, subtly shaping judgments without deliberate adjustment (Jacowitz & Kahneman, 1995; Wilson et al., 1996). While both models were influential, they are now seen as too narrow to explain the full range of anchoring effects. More recent research has given more weight to the Selective Accessibility Model (Mussweiler & Strack, 1999a, 1999c), which involves an initial phase of hypothesis-consistent testing—retrieving information that supports the anchor's plausibility—followed by a second phase shaped by that primed knowledge. Specifically, evaluators at the comparative judgment stage consider the proposition of whether the anchor value is a plausible answer by mentally testing it, a process called "confirmatory search" (Chapman & Johnson, 1999), thereby activating mental versions of the concept that are consistent with the anchor value. This activation makes values of the concept that



are proximate to the anchor more cognitively available. This priming of anchor-proximate values disproportionately influences the estimated value at the absolute judgment stage, thereby shifting the final judgment toward the anchor value. To summarize: "judgmental anchoring constitutes a phenomenon where, in spite of active denial, merely comparing a target with an arbitrary standard fosters judges' beliefs that the target's true value approximates the standard, so that here, *comparing is believing*" (Mussweiler & Strack, 1999a, p. 162). In line with this theory, anchoring effects were found to be quite concept-specific, in the sense that they influence the estimated value of the particular concept considered in conjunction with the anchor, rather than prompting anchor-proximate values of other concepts. For example, an anchor that pertains to the height of the Brandenburg Gate had an effect on estimates of the height of this gate, but much less influence on estimates of its width (Strack & Mussweiler, 1997). A more recent theory has suggested that anchoring effects arise not only from shifts in belief about the target concept but from a momentary adaptation of the response scale to the anchor (Frederick & Mochon, 2012); however, such scale distortions were not replicated in recent trials (Bahník, 2021). Our findings weigh in on this debate by showing that anchoring effects are both concept-selective and non-momentary, in line with the Selective Accessibility Model.

Much of the abundant literature on anchoring focuses on its detrimental implications for judgment and decision making in economic (e.g., Belsky & Gilovich, 2009; Green et al., 1998), political (e.g., Arceneaux & Nicholson, 2024) and even legal and moral choices (e.g., Statman et al., 2020). In social science methodology, anchoring is also considered a potential source of bias in survey and experimental designs, as



unintended anchors may unduly influence responses (Cavallo et al., 2017; Haaland et al., 2023). However, in the following section we present a simple experimental design that adopts anchoring as a feature rather than a bug, incorporating it deliberately for the purpose of treating beliefs in behavioral experiments.

*Anchoring-based treatment*

To study the effect of a belief regarding a certain concept $A$ on a dependent variable $Y$ (e.g., a decision or behavior), we randomly assign participants to an anchoring procedure with either a high or low value $a$, and present the following two generic questions:

(1) Do you think the value of "$A$" is greater than or less than "$a$"? [greater / less]

(2) What is the value of "$A$" in your opinion? [an open numeric response].

Question (1) prompts respondents to consider a low or high value of the concept $A$, thereby treating their estimated value of this concept using the anchoring effect. To the extent that anchoring is effective, randomizing the value of $a$ in question (1) yields two groups that differ only in their expected beliefs about the value of $A$. Next, respondents answer question (2), thus providing a posttreatment measure of the belief. A comparison of the answers of participants exposed to the high and low anchor values can be used to assess the effectiveness of the anchoring-based treatment (a manipulation check), or as the first-stage of an IV estimation (as detailed below). This procedure is followed (immediately or not) by a reported or behavioral measurement of the dependent variable(s). This simple design, which corresponds to the classic two-stage anchoring



experiment, offers the possibility of converting an otherwise observational study into an experiment by adding just one simple question.

*Integrating the anchoring-based treatment into instrumental variable estimation*

Estimating the causal relationship between the belief ($belief$) and the outcome ($Y$) can be obtained using ordinary least squares (OLS) regression, under the assumption that the belief is uncorrelated with the error term $\varepsilon$:

$$Y = \beta_0 + \beta_1 belief + \varepsilon. \tag{1}$$

However, since the belief is *endogenous*, the error term in the model, $\varepsilon$, is statistically dependent on the belief due to a set of covariates ($C_1, \ldots, C_K$) not included in the model, which may bias the estimator of the causal effect, $\beta_1$. The true model that includes these covariates is:

$$Y = \beta_0 + \beta_1 belief + \gamma_1 C_1 + \cdots + \gamma_K C_K + \varepsilon. \tag{2}$$

Only if all covariates are included in the estimated regression, can the coefficient $\beta_1$ be interpreted as the causal effect of the belief on the outcome. This requirement is rarely met, either because some covariates are unobserved, or because they are unknown. The bias caused by omitted variables can be positive or negative and of any magnitude, making the regression model with omitted variables unreliable for causal estimation purposes.



An established econometric technique for addressing omitted variable bias is instrumental variable (IV) estimation.[1] An instrument is a variable that influences $Y$ only through its effect on the endogenous variable in question (in our case, the belief). This assumption, known as the exclusion restriction, implies that the instrument is not included in Equation (2).

The random assignment of respondents to different anchor values satisfies the exclusion restriction, thus provides an IV for estimating the causal effect of the endogenous belief in question on the dependent variable. For each participant $i$, we have a binary variable ($high\_anchor_i$) that takes the value 1 for participants assigned to the high anchor value and 0 for those assigned to the low anchor value, a posttreatment valuation of the belief ($belief_i$), and a measure of the outcome decision ($Y_i$). The estimation procedure can be conveniently represented as a two-stage least squares (2SLS) model. The first-stage equation regresses the endogenous belief on the anchor (instrument):

$$belief_i = \alpha_0 + \alpha_1 high\_anchor_i + \epsilon_i. \tag{3}$$

The instrument is useful only if it has a sufficiently strong effect ($\alpha_1$) on the endogenous belief ($F > 10$; Sovey & Green, 2011). The first-stage regression provides predicted values of the belief ($\widehat{belief}_i$). Using these predicted values in the second-stage regression (instead of the endogenous ones) results in an unbiased and consistent estimate of the effect ($\beta_1$) of the belief on $Y$:

---

[1] For more details on IV estimation, see Angrist and Pischke (2009, Chapter 4) and Sovey and Green (2011).



$$Y_i = \beta_0 + \beta_1 \widehat{belief}_i + \varepsilon_i, \tag{4}$$

Finding a valid IV is often difficult, and is the main obstacle in applying this method. ABCD satisfies the conditions required for IV estimation. The randomly assigned anchor value is expected to have a causal effect on the belief, namely the anchoring effect. While the exclusion restriction assumption cannot be empirically verified (Sovey & Green, 2011), it is highly plausible that the only causal path between the IV (the anchoring-based treatment) and the outcome variable is via the endogenous belief. The anchoring-based treatment is essentially an uninformative prompt regarding the endogenous belief, and the cognitive mechanism of selective accessibility (Strack & Mussweiler, 1997) that underlies anchoring implies that considering an anchor influences the accessibility of anchor-proximate values of the target concept but does not similarly affect other concepts. For these reasons, we consider the exclusion restriction assumption to be highly plausible in the context of ABCD.

In summary, ABCD has several advantages. First, it offers the required conditions for careful causal inference by using the assignments of the anchor as an IV to estimate the effect of a belief on an outcome variable. Second, the anchoring-based treatment avoids many of the ethical challenges that pervade information-provision experiments. Asking respondents to merely consider a possible value of a concept in the comparative judgment stage is ethically unproblematic. Such tasks are common in surveys and other situations in everyday life (e.g., group deliberation), and most importantly, respondents are not called upon to adopt a particular concept-related value. Admittedly, the use of anchoring to treat beliefs involves a non-transparent manipulation; but it is one of the subtlest types of experimental manipulation. Third, the commonly raised concern about



whether priming effects operate by selectively activating the intended mental concept (Cohn & Maréchal, 2016) is addressed by the explicit posttreatment measure of belief.

Fourth, ABCD avoids biases that are related to source effects, which is a notable advantage over information provision experiments. Studies in persuasion and communication have shown that various source attributes, such as credibility, trustworthiness, and similarity, influence the way people process, evaluate and accept messages (e.g., Bettinghaus et al., 1970; Chaiken, 1980; Chen, 1999; Petty et al., 1986). Source effects in the context of information provision experiments may act in several ways: (1) They may alter the weight allotted to particular information. Such bias may stem from increased weight due to the source being an expert or otherwise influential. Conversely, if the source is deemed unfavorable, this bias may act to diminish the effect of the information; (2) Attitudes related to the source may affect the outcome other than through a change in the information, thereby violating the exclusion restriction assumption (Stantcheva, 2023). These potential problems are addressed by ABCD, as it requires no mention of a source, and the treatment is administered through self-reflection, since the anchor value is simply presented as a possible (yet not necessarily correct) option to consider. This method is also expected to reduce the motivation to infer a source, thus further decreasing the potential for biases due to source-related attitudes. Lastly, as shown below, the anchoring effect is concept-specific and persists long enough (several days) to suggest that it is not merely a momentary adaptation in response scale.

**Empirical assessment of ABCD**

To assess the utility of ABCD, we evaluate four attributes of this design. First, we study the effectiveness of anchoring-based treatments in replicating two established causal



relationships. Second, we estimate the durability (and decay) of anchoring effects in follow-up studies. Third, we further explore the relationship between anchor values and the strength of anchoring treatments. Lastly, using placebo tests, we assess the selectiveness of anchoring-based treatments, namely, the extent to which they affect the intended belief without influencing other beliefs.

For these purposes, we conducted eight experimental studies ($N = 3,299$) in four rounds, as outlined in Table 1. The experiments were IRB approved, and four of them were preregistered. Round one (July 1–4, 2024; $n = 1,012$) included two substantively different experiments.[2] Round two (July 23–30, 2024; $n = 1257$) included modified replications of the two experiments through the addition of a "no anchor" condition in both experiments and revising the anchor values in one of them. Round three (August 4–6, 2024; $n = 895$) is a follow-up study that began five days after the end of round two, to assess the durability of the anchoring effect among respondents who participated in round two. The fourth round of experiments replicated the two experiments with multivalued anchors to explore the relationship between anchor values and IV estimation effectiveness. In addition, we utilized the fact that rounds one and two included three and two experiments, respectively, in the same online survey, to assess the selectiveness of anchoring-based treatments by conducting a set of placebo tests. Data and code are available at: https://osf.io/s4ke6/?view_only=973e746c19fa46a49858b34d42b3c52d.

---

[2] This round included also a third experiment, aimed to address a distinct substantive research question using ABCD: the effect of individual beliefs about the effectiveness of the flu vaccine on the intention to get vaccinated. This question had not yet been tested experimentally, and previous observational studies suggest that such beliefs are not a primary factor in vaccination behavior (Blank et al., 2009; Santos et al., 2017).



Table 1: Overview of the studies

| | Round 1 July 1-4, 2024 n = 1012 | Round 2 July 23-30, 2024 n = 1257 | Round 3 Aug. 4-6, 2024 n = 895 | Round 4 Sep. 1-7, 2024 n = 1030 | Placebo tests |
|---|---|---|---|---|---|
| **Motivation** | Assessing ABCD | | Durability of anchoring treatments | Strength of anchoring treatments | Assessing the selectiveness of anchoring-based treatments |
| **Domain I** *Macroeconomic beliefs and personal economic expectations* | Anchor values based on Roth and Wohlfart (2020) | Replication with revised anchor values and the addition of a 'no-anchor' condition | Follow-up study with participants of round 2, to measure lagged anchoring-effect | Replication with 51 anchor values | Multiple regression estimates of related and unrelated anchoring effects |
| **Domain II** *Perceived social norm of charitable donations and willingness to donate* | Anchor values that are multiples of 12 and differ in their order of magnitude | Replication with the addition of a 'no-anchor' condition | Follow-up study with participants of round 2, to measure lagged anchoring-effect | Replication with 49 anchor values | |

*First round of experiments*

The experiments in the first round were preregistered.[3] A total of 1,012 respondents, recruited by iPanel (a survey company conducting online surveys in Israel) participated in three survey experiments, two of which are reported in this paper. Respondents first provided basic demographic details,[4] and then completed the ABCD experiments (the full questionnaire is provided in Section I of the Supplementary Information).

One experiment tested the effect of macroeconomic beliefs on personal economic expectations, in the spirit of the information provision experiment conducted by Roth and Wohlfart (2020). As in the original study, respondents first read an explanation of the meaning of recession. This was followed by an anchoring-based treatment of the likelihood of a recession, which replaced the experts' predictions used in the original study. Respondents were randomly assigned to one of two conditions, by being asked to consider one of two alternative probabilities, 5% or 40%, in the following question:[5]

---

[3] https://aspredicted.org/NSP_4M7.
[4] The sample consisted of 511 women, 501 men; age groups: 2 (<25); 187 (25-34), 173 (35-44), 200 (45-54), 204 (55-64), 172 (65-74), 74 (≥75).

In your opinion, is the likelihood of an economic recession in the third quarter of 2024 (during July–September) higher or lower than 5% (40%)?

o   Higher than 5% (40%)

o   Lower than 5% (40%)

Next, respondents were asked to estimate the likelihood of a recession in the next quarter, by setting a slider between 0 and 100%. This response provided a measure of the endogenous belief, and was followed by four outcome variables: (1) expectations regarding unemployment level on an ordinal scale;[6] (2) expectations regarding unemployment level on a percentage scale; (3) expectations regarding household financial prospects on an ordinal scale;[7] (4) expectations regarding employment status level on a percentage scale.[8] To minimize carryover effects on subsequent experiments, we had the respondents answer a filler question after completing each experiment. For one filler question they were asked to rate three diseases from most dangerous to least dangerous, and for the other they were asked to rate five NGO's based on the importance of their social activity.

---

[5] In the original Roth and Wohlfart (2020) study, the information treatments (expert's predictions) referred to 5% or 35% probabilities.
[6] Unemployment will strongly increase, somewhat increase, remain the same, somewhat decrease, strongly decrease.
[7] Much worse off, worse off, somewhat worse off, about the same, somewhat better off, better off, much better off.
[8] These are four outcome variables that were affected by the treatment of macroeconomic beliefs in Roth and Wohlfart's (2020) study. Six additional outcome variables were included in their study, but only two of them – county unemployment (percent) and firm profits (ordinal) – were influenced.

The second experiment in the first round tested the effect of individuals' beliefs about the descriptive social norm of charitable donations on the intention to donate. Previous observational and experimental studies have shown that beliefs about a descriptive social norm – how most people actually behave – affect a range of (reported and actual) behaviors (Cialdini, 2007), including charitable donations (for a review, see van Teunenbroek et al., 2020). We utilized ABCD to test this effect. The experiment began with an anchoring-based treatment of the descriptive social norm of charitable donations, where respondents were asked to evaluate one of two values of a mean yearly household donation:

In your opinion, is the mean yearly household donation to NGOs and social causes higher or lower than 120 NIS (12,000 NIS)?

o   Higher than 120 NIS (12,000 NIS)

o   Lower than 120 NIS (12,000 NIS)

Since the estimated value addresses a yearly period, we opted for anchors that are clear multiples of 12. Additionally, to account for the fact that humans evaluate numeric scales logarithmically rather than linearly (Dehaene, 2003), the anchor values differ in their order of magnitude ($12{\times}10^1$ and $12{\times}10^3$).



Following this question, respondents were asked to estimate the value of the mean yearly household donation by writing the number.[9] In this particular case, the treated value does not have an upper bound (unlike recession expectations that range between 0 and 100%).[10] Lastly, they were asked to indicate the sum of money they would donate to an NGO or a social cause of their choice, if they were to win a monetary prize of 100,000 NIS.[11]

*Results of the first round of experiments*

The descriptive results of treating recession expectations and average yearly household donation (on a logarithmic scale) are presented in Figures 1A and 1B, respectively. The regression results of the two experiments are reported in Table 2. We begin our analyses by estimating the correlations between the treated beliefs (recession expectations and descriptive social norm) and the corresponding outcome variables, as given in Equation (1). These correlations, represented by the OLS estimate of $\beta_1$, are reported for each of the four economic and the single donation outcome variables. As there are likely important covariates that are omitted from the regression model (1), the estimate of $\beta_1$ should not be interpreted as the causal effect of the belief on the outcome. To identify this causal effect, we instrument the belief using the assigned anchor values, as given in Equations (3) and (4).

---

[9] The software (Qualtrics) was programmed to accept only a numeric response that exceeds or equals to 0.
[10] Responses of above 100,000 NIS (~26,400 US$) were excluded. These included eleven extreme responses (1.1%).
[11] An acceptable answer was only a numeric response between 0 and 100,000.



The correlations between recession expectations and the outcome variables are statistically significant and in the theoretically expected direction (note that higher ordinal values reflect greater optimism). However, as explained above, these correlations cannot be interpreted as causal relations. The anchoring effect[12] on recession expectations is statistically significant yet small and insufficient to conduct IV estimation (3.49 percentage points (pp), $p = 0.026$, $F = 4.96$). Our attempt to replicate Roth and Wohlfart's (2020) study led us to use similar anchor values to those used in their original information provision experiment. In retrospect, this decision proved misguided, given the level of economic pessimism during the time and place of our experiment (Israel, July 2024). As shown in Figure 1A, both anchor values of recession expectations (solid red and blue lines) are lower than the mean belief regarding recession likelihood. These results correspond to data collected routinely by the Israeli Central Bureau of Statistics, which indicate that our experiment took place during one of the least economically optimistic periods since January 2021 (see Section II in the Supplementary Information). These anchor values are clearly too low, leaving us with only correlational results that substantively replicate those of Roth and Wohlfart (2020).

The correlation between the estimated social norm of donation and the willingness to donate is positive and statistically significant, as theoretically predicted. The descriptive results of treating the social norm of donation (estimated average yearly household donation) are presented in Figure 1B. The anchoring effect on this log-transformed estimate is statistically significant and sufficiently large (0.744 log(NIS), $p <$

---

[12] The anchoring effect is defined as the difference between the mean response in the high-anchor group and that in the low-anchor group.



0.001, $F = 321.33$). Estimating the effect of this belief on the willingness to donate in the IV analysis yields a significant positive effect. Multiplying the estimated yearly household donation by 10 increases the intended donation by a factor of 1.92. The difference between the OLS and IV coefficient estimates indicates the importance of correcting for omitted covariates in accurately estimating the causal effect.

Table 2: Main results of the first round of experiments

| | Economic expectations | | | | Philanthropic behavior |
|---|---|---|---|---|---|
| | National unemployment (ordinal) | National unemployment (percent) | Household finances (ordinal) | Personal unemployment (percent) | Logged intended donation |
| **OLS** | | | | | |
| Posttreatment belief | -0.013*** (0.001) | 0.612*** (0.027) | -0.013*** (0.001) | 0.260*** (0.030) | 0.369*** (0.046) |
| **IV** | | | | | |
| Posttreatment belief | - | - | - | - | 0.283** (0.093) |
| Anchoring effect | | 3.49 pp* | | | 0.744*** |
| 1st stage $F$-statistic | | 4.96 | | | 321.33 |
| **Descriptive statistics** | | | | | |
| Observations | | 1012 | | | 1003 |
| Mean outcome | 2.25 | 48.86 | 3.56 | 20.80 | 3.198 |
| SD outcome | 0.833 | 26.24 | 1.085 | 24.72 | 1.126 |

The table shows OLS and IV coefficient estimates based on regression of the outcome variables on the beliefs, the first-stage anchoring effect with the corresponding F-test, and basic descriptive statistics. The anchoring effect in the economic expectations experiment did not pass the F>10 criterion, and hence the IV estimates were not calculated. Standard errors are in parentheses. Significant at *5%, **1%, and ***0.1%.

In summary, out of the two studies, we successfully replicated one – the descriptive social norm and donation – by applying ABCD, and we failed to replicate the economic expectations experiment due to anchor values that were not appropriate for the baseline mean belief. In the second round of experiments, we replicated these two studies



by modifying the anchor values in the economic expectations experiment and adding a "no-anchor" condition in both studies.

Figure 1: Anchoring effects on beliefs.

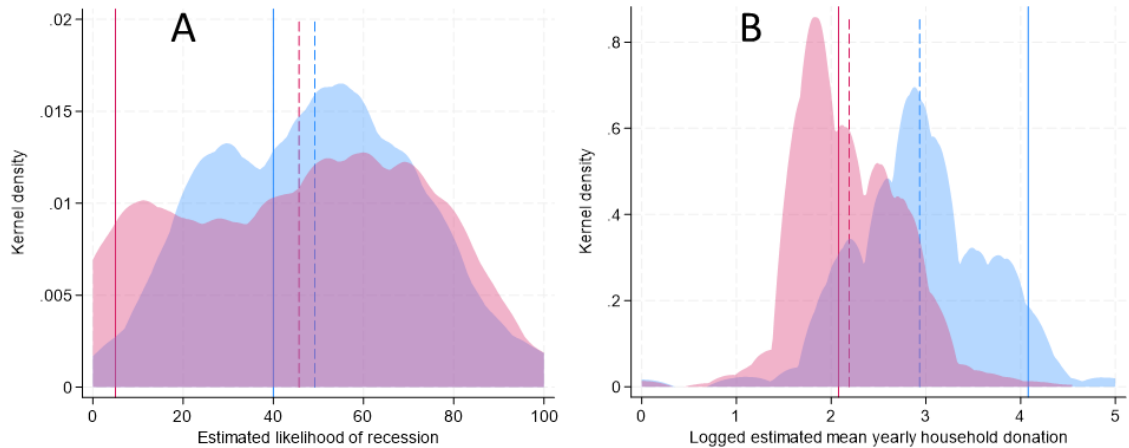

Each panel presents kernel smoothed distributions of beliefs under low (red) and high (blue) anchor values. Solid blue and red lines indicate high and low anchor values, respectively. Dashed blue and red lines indicate the overall mean belief under high and low anchor values, respectively.

*Second round of experiments*

The second round of experiments was conducted among a sample of 1,257 respondents, recruited by iPanel, who did not participate in the first round.[13] The experimental procedure was identical to that of the first round, with the following changes: (1) based on the distribution of estimated likelihood of recession in the first round, the new anchor values were set to 10% and 80%, which roughly correspond to the 10th and 90th percentiles; (2) each of the two experiments included a third, no-anchor condition.[14]

---

[13] The sample consisted of 606 women, 650 men, and 1 other gender; age groups: 1 (<25); 241 (25-34), 244 (35-44), 209 (45-54), 214 (55-64), 222 (65-74), 126 (≥75).

[14] The sample size of the no-anchor condition in the economic experiment (n = 245) was intentionally smaller than those of the anchor conditions, as priority was given to reserving sufficient sample sizes for the revised anchor conditions.



The descriptive results of the treated and untreated recession expectations and the estimated social norm are presented in Figures 2A and 2B, respectively. The regression results of the two experiments are reported in Table 3. The correlations between the treated beliefs (recession expectations and descriptive social norm) and the corresponding outcome variables, are statistically significant and substantively similar to the previous round of this experiment.

As shown in Figure 2A, the revised anchor values for recession expectations (solid red and blue lines) are clearly below and above the untreated mean estimate (dashed green line). The anchoring effect is statistically significant and sufficiently strong to conduct an IV estimation (16.09 pp, $p < 0.001$, $F = 50.06$).[15] The new anchor values resulted in a 4.6-fold larger effect, compared with the first round. Estimating the effect of this belief on the four dependent variables yielded mixed results. A significant positive effect was found on the estimated likelihood of an increase in national and personal unemployment. An increase of 1 pp in the likelihood of recession increases expected national and personal unemployment by 0.537 pp and 0.229 pp, respectively. No significant effects were found on the ordinal measures of expected national unemployment and on household finances.

Despite the fact that using linear regression is not appropriate for ordinal outcome variables, we conducted these analyses for the sake of replication. To address this issue, we converted the two ordinal variables to binary variables, choosing cut-off values that are closest to the medians. Estimating the causal effect of recession expectations on these binary variables resulted in a statistically insignificant effect on national unemployment prospects ($\beta$=0.002, $p$=0.316), and a statistically significant and small positive effect on household financial prospects ($\beta$=0.004, $p$=0.049). The latter effect is in the inverse direction, and therefore does not replicate the original result of Roth and Wohlfart (2020).

---

[15] IV estimates in both studies in the second round used the two conditions – 'high anchor value' and 'no anchor value' – as instruments.

As Figure 2B shows, the addition of untreated belief regarding the average yearly household donation confirms that the anchor values are below and above the untreated mean belief. As in the previous round, the anchoring effect on this belief is statistically significant and sufficiently large (0.834, $p < 0.001$, $F = 173.51$). Estimating the effect of this belief on the willingness to donate in the IV regression results in a significant positive effect. Multiplying the estimated yearly household donation by 10 increases the intended donation by a factor of 1.61. This effect is 16% smaller than the effect found in the previous round.



Table 3: Main results of the second round of experiments

| | Economic expectations | | | | Philanthropic behavior |
|---|---|---|---|---|---|
| | National unemployment (ordinal) | National unemployment (percent) | Household finances (ordinal) | Personal unemployment (percent) | Logged intended donation |
| **OLS** | | | | | |
| Posttreatment belief | -0.014*** (.001) | 0.609*** (0.023) | -0.013*** (0.001) | 0.224*** (0.025) | 0.438*** (0.045) |
| **IV** | | | | | |
| Posttreatment belief | -0.0004 (0.003) | 0.537*** (0.084) | 0.003 (0.005) | 0.229* (0.092) | 0.194* (0.098) |
| Anchoring effect | 16.09 pp*** | | | | 0.834*** |
| 1ˢᵗ stage $F$-statistic | 50.06 | | | | 173.51 |
| **Descriptive statistics** | | | | | |
| Observations | 1257 | | | | 1232 |
| Mean outcome | 2.333 | 50.58 | 3.66 | 19.70 | 3.062 |
| SD outcome | 0.885 | 26.89 | 1.179 | 24.19 | 1.216 |

The table shows OLS and IV coefficient estimates based on regression of the outcome variables on the belief, the first-stage anchoring effect with the corresponding F-test, and basic descriptive statistics. Standard errors are in parentheses. Significant at ∗5%, ∗∗1%, and ∗∗∗0.1%.

To summarize the results so far, correcting the anchor values in the economic expectations experiment substantively increases the anchoring effect and facilitated the application of ABCD to test the effects of recession expectations. Our results largely replicate the results of Roth and Wohlfart (2020) with respect to expected national and personal unemployment, but do not find support for the effects on the ordinal measures of expected national unemployment and expected household finances. The different results may stem from the contextual differences between the two studies (2017 US vs. 2024 Israel), or possibly from source effects present in the information provision experiment but absent in the ABCD experiment. The replication of the philanthropic behavior experiment yields results closely aligned with those of round one. The anchoring effect



and the IV estimation are similar in the two rounds, and consistently support the hypothesis that belief about this norm positively affects willingness to donate. These results demonstrate the utility of ABCD and its robustness. However, they also point to the importance of specifying anchor values that maximize the first-stage anchoring effect, as we discuss in more detail below. Next, we present a follow-up study carried out several days after the second round of experiments to assess the rate of decay of the anchoring effects.

Figure 2: Anchoring effects on beliefs vs. untreated beliefs.

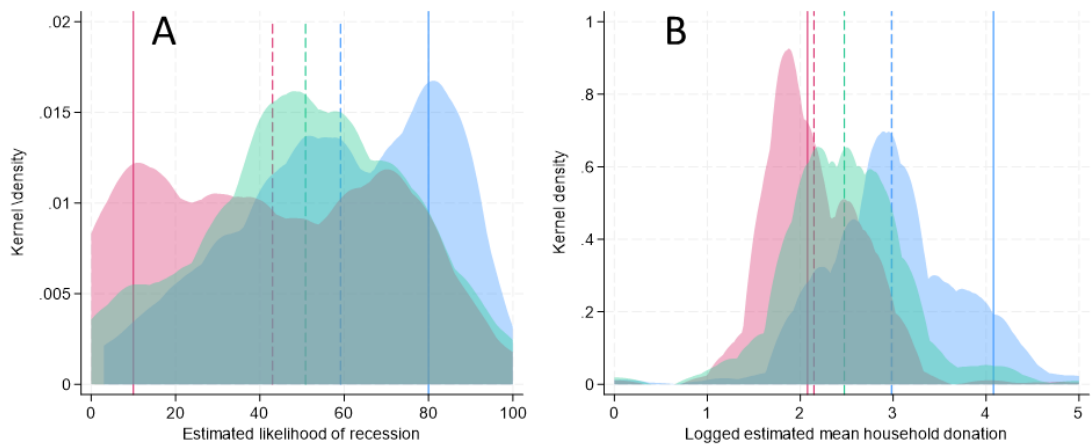

Each panel presents kernel smoothed distributions of beliefs under low (red) and high (blue) anchor values, and under no anchor (green). Solid blue and red lines indicate high and low anchor values, respectively. Dashed blue, red and green lines indicate the overall mean belief under high, low, and no anchor values, respectively.

*Durability of anchoring effects*

The few studies on the durability of anchoring effects provide mixed evidence. Mussweiler (2001) finds that such effects are detected with no apparent decay one week after exposure, while Blankenship et al. (2008) also detect week-old anchoring effects but report significant decay. Blankenship et al. (2008) also find that more thoughtful



consideration of the outcome judgment – induced by lower cognitive load – leads to more enduring anchoring effects.

To study the rate of decay of anchoring effects, we conducted two preregistered follow-up experiments,[16] which began five days after completion of the second round of experiments. Between August 4–6, we invited all participants of round two to take part in the same two experiments, administered in random order, this time without anchors. Of these, 895 participants (72.5%) completed the follow-up experiments.[17] For each participant we recorded the interview dates in the second and follow-up rounds, to create a measure of the time elapsed since treatment exposure (in days). The mean time lag was 8.8 days (SD = 2.099). Using participants' identification numbers in the survey company's records, we matched each respondent's follow-up beliefs (recession expectations and estimated yearly household donation) with their anchor values in the second round, and assessed the change in the effect of anchors over time.

The effects on both recession expectations and estimated yearly household donation are still detectable 8.8 days after anchor exposure (4.71 pp, $p = 0.010$, $F = 6.59$; and 0.321, $p < 0.001$, $F = 31.41$, respectively), but the lagged effects show a significant decay relative to the instantaneous effects. Figure 3 presents the instantaneous and lagged anchoring effects in the two experiments. More detailed temporal analyses show similar anchoring effects among participants whose repeated survey took place 5–9 days ($M =$

---

[16] https://aspredicted.org/see_one.php.
[17] The sample consisted of 420 women, 474 men, an 1 other gender; age groups: 145 (25-34), 183 (35-44), 154 (45-54), 174 (55-64), 152 (65-74), 87 (≥75).



6.96) and 10–14 days ($M = 10.78$) after anchor exposure, suggesting that the effect persists for at least two weeks (see Section IV in the Supplementary Information).

Figure 3: Anchoring effects decay.

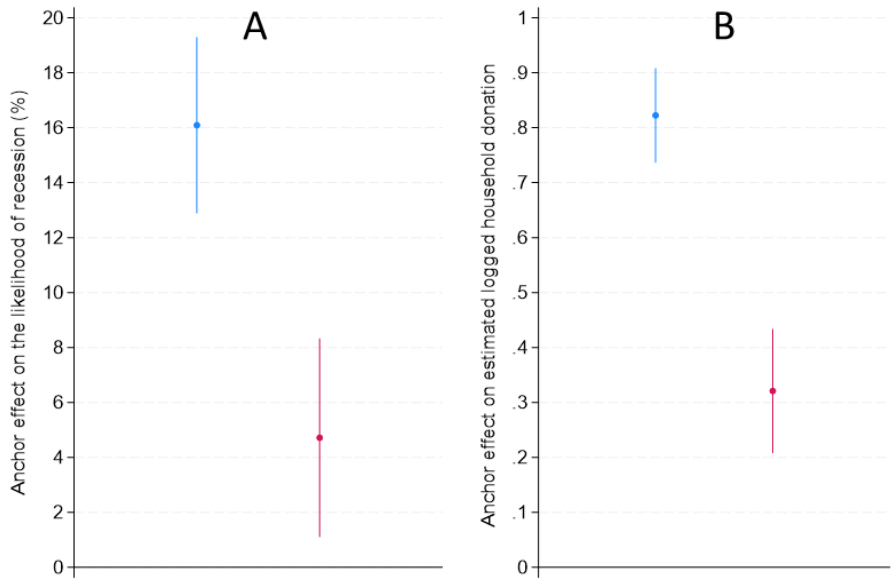

*Note*: Panels A and B present the instantaneous (blue) and lagged (5–14 days, $M = 8.8$) (red) anchoring effects on the recession expectations and estimated mean household donation, respectively, with 95% CIs.

These results suggest that while anchoring effects are quite durable – as they can be traced up to 9 days after exposure – they appear to decay substantively over this time period. This is consistent with the findings of Blankenship et al. (2008) but not with those of Mussweiler (2001). This temporary quality of anchoring-based treatments, together with their nondeceptive nature, jointly offers a relatively ethical method for treating people's beliefs. The following section delves deeper into the design question of determining the anchor values in order to maximize the anchoring effect.



*The relationship between anchor value and anchoring effect*

The first and second rounds of the economic expectations experiment demonstrated the critical role of anchor values in effectively treating baseline beliefs. To explore this issue more systematically, we replicated the two experiments with a sample of 1030 new respondents, replacing the two-anchor design with a set of multi-valued anchors. The economic experiment included 51 randomly assigned anchor values (from 0 to 100% in 2% intervals), and the donation experiment included 49 randomly assigned anchor values, from 0 to 1,000,000 NIS in an approximately logarithmic scale (see Section III in the supplementary information).

The effects of the anchor values on recession expectations and estimated mean household donation are presented in Figures 4A and 4B, respectively. Across the entire range of recession likelihood anchor values (0–100%), mean recession expectations varied between ~42% and ~63%. The relationship between the anchor value and the recession likelihood is not linear. Anchor values in the 0–40% range resulted in roughly the same mean likelihood estimates of about 43%. This pattern suggests that respondents were able to adjust their estimates away from the anchor the further its value deviated below 40%. These results are consistent with previous findings according to which anchor values that are closer to the baseline belief have stronger effects than distant (or implausible) anchor values (Mussweiler & Strack, 2001; Teovanović, 2019; Wegener et al., 2001). Respondents exposed to anchors in the 40–55% range were most influenced by the anchor value – likely because this range is relatively proximate to the mean baseline estimate (see Figure 2A). Another range in which the anchor was influential, albeit more moderately, was between 80–90%, before plateauing in the 90–100% range. These results



clarify (and substantively replicate) the results of the first and second rounds of this experiment. The anchor values of 5% and 40% used in the first round (dashed red and blue lines, respectively, in Figure 4A) fell within a range where respondents were relatively insensitive to the anchor. By contrast, the 10% and 80% anchor values used in the second round (solid red and blue lines, respectively, in Figure 4A) lie clearly below and above the range of greatest sensitivity, resulting in a sizable anchoring effect.

Across the range of yearly household donation values between zero and 1 million NIS, mean estimates varied between approximately 80 ($10^{1.9}$) and 1,300 ($10^{3.11}$) NIS. In this context as well, the relationship between anchor value and estimated household donation is not linear. A log-log linear relationship between anchor value and donation estimate holds for anchor values between 50 ($10^{1.7}$) and 20,000 ($10^{4.3}$) NIS. Again, the range of anchor sensitivity is located close to the baseline mean estimate (see Figure 2B). Below and above this anchor value range, we find a lack of sensitivity to anchor values, and even a reversal of the anchoring effect – mainly in the upper range, where respondents appear to increase their adjustment away from the anchor. These results clarify why the anchor values used in the first two rounds of this experiment (represented by the solid red and blue lines) were effective, as both anchors were located close to the edges of the anchor value range within which respondents were sensitive to the anchor value. However, the two anchors could have been set at more extreme values to increase the anchoring effect even further.

More specifically, for a binary IV, the estimator of $\beta_1$ in Equation (4) is given in a simple form known as the Wald estimator (Cameron, 2005, Section 4.8.4):



$$\beta_{IV} = \frac{\bar{Y}_2 - \bar{Y}_1}{\overline{belief}_2 - \overline{belief}_1}, \tag{5}$$

where the indices 1 and 2 of the averages in Equation (5) denote the low and high anchoring groups, respectively. The estimator $\beta_{IV}$ is unbiased and its variance depends on the anchoring effect, becoming smaller as the denominator of Equation (5) increases. Thus, the estimated variance is inversely proportional to the squared difference in mean beliefs between the two anchoring groups: $\left(\overline{belief}_2 - \overline{belief}_1\right)^2$. An optimal design chooses anchor values that maximize the difference of the average beliefs.

To demonstrate the idea, consider the estimated smoothing curve for the donation experiment in Figure 5B. For the anchors used in the first and second rounds – 120 and 12,000 NIS, or 2.08 and 4.08 on a log scale – the mean beliefs ($\overline{belief}_1$, $\overline{belief}_2$) are 145 and 1,227 NIS, or 2.16 and 3.09 on the log scale, respectively, with a difference of 0.93 on the log-scale. The maximum difference among the average logged-beliefs in the experiment is 1.34 on the log-scale, obtained for anchors 10 and 40,000 NIS. Thus, using the latter anchors could decrease the variance by about $0.5 = (0.93/1.34)^2$.



Figure 4: Anchoring effects across anchor values in the two experiments.

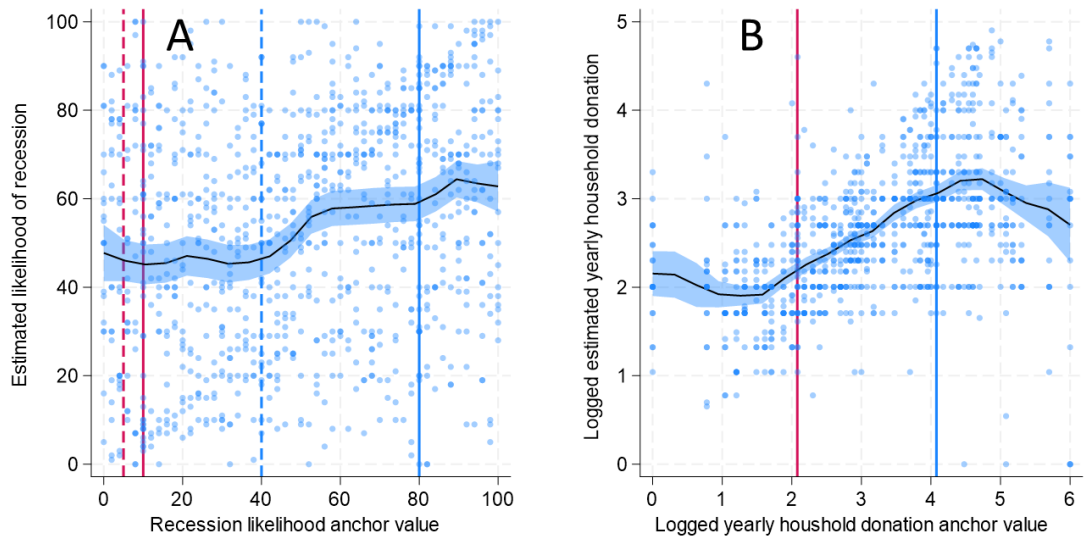

*Notes*: Panel A presents the effects of anchors between 0 and 100 precent on the estimated likelihood of recession. Each blue dot represents a participant, and the smoothed black line represents local polynomial, with 95% CIs. The dashed red and blue lines mark the 5% and 40% anchor values, respectively, and the solid red and blue lines indicate the 10% and 80% anchor values, respectively. Panel B presents the same information for the effects of logged anchor values between 0 and $10^6$ on the logged estimated mean household donation. The solid red and blue lines indicate the 120 and 12000 anchor values, respectively.

*The selectiveness of the anchoring-based treatment*

The selectiveness of the anchoring-based treatment – namely, its effect on the intended belief only – carries important implications for its applicability. This quality relates to the plausibility of the exclusion restriction assumption, which pertains to the validity of the causal inference. Additionally, treatment selectiveness also relates to the possibility of using several anchoring-based treatments, each addressing a different belief, within a single task.

As noted above, some previous evidence suggests that anchors are concept-selective (Strack & Mussweiler, 1997). We take advantage of the fact that rounds one and



two included three[18] and two experiments, respectively, in the same online survey to conduct a set of placebo tests. The order of experiments was fixed, with recession expectations first, vaccine effectiveness second (in the first round), and estimated mean household donation last (with filler questions between experiments). To assess anchor selectiveness, we use placebo tests, by fitting OLS regressions to estimate all the possible anchoring effects within each round on the posttreatment beliefs. Any significant unintended effect – i.e., of an anchoring-based treatment on an unrelated belief (e.g., the effect of an anchoring-based treatment of recession likelihood on estimated mean household donations) – provides evidence that challenges the assumption of anchor selectiveness.

Table 4 presents the results of these analyses (coefficients of placebo tests are presented bold font). The first regression (left column) estimates the effects of the three anchoring-based treatments from round one on the estimated mean household donation. These results show no effects of unrelated treatments and a significant effect of the relevant one. Next, we estimate the effects of the two anchoring-based treatments on the estimated vaccine effectiveness (middle column). This analysis also shows a significant effect of the relevant treatment only. Lastly, we assess the effects of the anchoring-based treatments in round two on the estimated mean household donation (right column). Note that in the second round we included 'no anchor' conditions in the two experiments, and the analysis includes them as well, with the 'low-anchor' conditions serving as reference.

---

[18] Although the vaccine effectiveness experiment is not an integral part of this research, the use of an anchor treatments in this experiment offers the opportunity to include it in this analysis, as it focuses only on first stages (the anchoring effects).



The only significant effects are those of the relevant treatments. These results provide

consistent evidence for the selectiveness of anchoring-based treatments.

Table 4: Selectiveness of anchoring-based treatments.

| | Round 1 | | Round 2 |
|---|---|---|---|
| | Logged estimated household donation | Estimating vaccine effectiveness | Logged estimated household donation |
| High recession expectations anchor | **-0.009** **(0.042)** | **-0.774** **(1.562)** | **-0.042** **(0.040)** |
| No recession expectations anchor | | | **0.009** **(0.050)** |
| High vaccine effectiveness anchor | **0.018** **(0.042)** | 14.35*** (1.563) | |
| High household donation anchor | 0.744*** (0.042) | | 0.826*** (0.044) |
| No household donation anchor | | | 0.320*** (0.044) |
| Constant | 2.187*** (0.041) | 48.24*** (1.344) | 2.157*** (0.039) |
| Observations | 1,003 | 1,012 | 1226 |
| Adj. R-squared | 0.241 | 0.075 | 0.229 |

Standard errors in parentheses; *** p<0.001, ** p<0.01, * p<0.5. Coefficients of placebo tests in bold.

**Discussion**

This paper introduces and empirically demonstrates the Anchoring-Based Causal Design

(ABCD), a novel experimental method designed to treat beliefs and estimate their causal

effects on outcomes, while overcoming significant limitations of existing approaches. A

central challenge in studying the effects of beliefs on decisions and behaviors is the

presence of omitted variable bias and confounding factors. While information provision

experiments attempt to address this by randomly treating beliefs, they often face ethical

concerns due to deception and inferential threats stemming from information source



effects. ABCD addresses these issues by combining the cognitive process of anchoring with instrumental variable (IV) estimation.

The core of ABCD involves a two-stage procedure: first, participants are asked a comparative judgment question, where they consider a randomly assigned high or low value in relation to a particular attribute of a concept (the anchor), followed by an absolute judgment task in which they provide their estimate for the concept's value (their expressed belief). The anchoring-based treatment is relatively short, especially when compared to typical information provision treatments. The randomly assigned anchor value serves as an instrumental variable that enables unbiased and consistent estimation of the causal effect of the belief on the outcome variable. A key strength of ABCD lies in its ability to avoid deception, as respondents are merely presented with a value to reflect upon, and are not asked to adopt a particular belief. Furthermore, it eliminates information source effects, since no external source is mentioned, and the belief treatment occurs through self-reflection. ABCD effectively addresses concerns about whether the treatment selectively influences the intended concept (Cohn & Maréchal, 2016) by incorporating an explicit measure of the posttreatment belief. Importantly, as suggested by the Selective Accessibility Model (Mussweiler & Strack, 1999b, 1999c; Strack & Mussweiler, 1997), our empirical results show that the anchoring effects are concept-specific, thereby increasing the plausibility of exclusion restriction assumption.

The utility of ABCD was evaluated across eight experimental studies. We assessed the effectiveness of anchoring-based treatments, replicated causal relationships, estimated the durability of anchoring effects, and examined the selectiveness of treatments. The household donation experiment consistently demonstrated an effective



anchoring-based treatment, successfully replicating the finding that beliefs about charitable donation norms positively affect willingness to donate across two rounds. By contrast, the first economic expectations experiment suffered from anchor values that were too low for the prevailing economic sentiment, resulting in an insufficient anchoring effect for valid IV estimation. However, after we adjusted the anchor values in the second round (from 5% / 40% to 10% / 80%), the anchoring effect significantly increased, enabling successful replication of the causal effect of recession expectations on expected national and personal unemployment. Our results show that while anchoring effects are detectable after an average of 8.8 days, they are significantly smaller compared to the immediate effect. These results align with Blankenship et al. (2008), who reported significant decay of week-old anchoring effects, as opposed to Mussweiler (2001), who observed no apparent decay. Finally, the analysis confirmed the selectiveness of anchoring-based treatments, by showing no spillover effects on unrelated beliefs, which bolsters the plausibility of the exclusion restriction assumption.

In view of the current methodological literature, ABCD challenges some of the assumptions that guide existing typologies of experimental methods. ABCD constitutes a priming experimental method for treating beliefs. As such, it extends the two types of established priming experiment – *conceptual priming*, which primes mental concepts (e.g., national identity), and *mindset priming*, which primes a way of thinking (e.g., risk-aversion) (Bargh & Chartrand, 2014) – and extends the typology of priming experiments provided by Stantcheva (2023). In other words, ABCD departs from the convention that priming experiments are restricted to treating concepts and mindsets, whereas information provision experiments are reserved for treating beliefs.



Our empirical analyses also contribute to the theoretical debate on the mechanism of anchoring. Specifically, our findings conform to the Selective Accessibility Model (Mussweiler & Strack, 1999b, 1999a; Strack & Mussweiler, 1997) and contradict the Scale Distortion Theory (Frederick & Mochon, 2012). The anchoring effects in both studies were not momentary, and were concept-specific as no spillover effect was detected. Still, further studies should be mindful of the theoretical possibility that, for specific concepts, the anchoring effect may operate in the short term by distorting the response scale rather than by priming the perceived value of the concept.

Despite its advantages, ABCD has several limitations. First, the effectiveness of the anchoring-based treatment is dependent on the choice of anchor values. The recession expectation experiment in the first round demonstrated that ineffective anchor values, i.e., those not appropriately aligned with the baseline mean belief, can result in a weak or insufficient first-stage effect for valid IV estimation. In order to choose effective anchor values, it is advisable to gather preliminary information on the baseline distribution of the belief to be treated, by conducting a pre-experimental pilot. Generalizing from the results of the two multivalued anchor experiments on recession expectations and household donations (see Section V in the Supplementary Information), we conjecture that anchor values that would approximately maximize anchoring effects are those of the 5th and 95th percentiles of the baseline distribution.

Another limitation is the decay of anchoring effects over time. While anchoring effects are detectable for about nine days, their substantive decay limits the direct applicability of ABCD for studies requiring long-term belief manipulation. Also, while ABCD replicated some of the results in the recession expectations experiment, it did not



fully replicate the effects on all outcome variables from the original Roth and Wohlfart (2020) study, particularly for ordinal outcome measures. These discrepancies could be due to contextual differences (e.g., the US economy in 2017 vs. the Israel economy in 2024) or, notably, the absence of source effects that were present in the original information-provision design. Furthermore, the use of linear regression for ordinal outcome variables, though done for replication purposes, is methodologically inappropriate. Finally, while subtle and nondeceptive, the anchoring-based treatment still involves a nontransparent manipulation, which may warrant further consideration depending on the research context.

The promising results of ABCD open several avenues for future research. Further research is needed to develop more systematic and efficient methods for determining optimal anchor values that maximize the anchoring effect. This could involve pretesting with multivalued anchors or developing adaptive experimental designs to identify sensitivity ranges. Second, a deeper investigation into the factors influencing the decay rate of anchoring effects could help in designing interventions that achieve more durable belief treatments where needed. Understanding the underlying cognitive processes that contribute to decay could inform strategies to sustain anchoring effects or develop methods for re-treating beliefs over time. Third, further exploration of the applicability of ABCD to a wider array of beliefs and behaviors in various social science domains (e.g., political science, public health, consumer behavior) and in cross-cultural contexts will clarify the range and boundaries of its utility.

In conclusion, we believe that ABCD represents a significant methodological advancement, offering a robust and ethically sound alternative for estimating the causal



effects of beliefs in experimental social science. Its ability to minimize confounding

biases while maintaining causal identification makes it a valuable tool for future research.

# Supplementary Information

## I.      Round-one questionnaire

The following survey is intended to learn about the opinions of different people regarding a number of topics in the fields of health, society, and consumerism. The time it takes complete the questionnaire is about 8 minutes.

You are not obligated to participate in this research, and you have the right to discontinue your participation without explanation at any stage during the completion of the questionnaire. The information is collected for academic research by researchers from the university. Payment is for completed questionnaires only. Your answers are anonymous and are not linked to information that may personally identify you. If you have a question regarding the research, you can contact [information omitted for blind review purposes]. By selecting "yes", I confirm that I understand the purpose of this research and how my information will be used.

**Consent**
o Yes
o No

Page Break

**Gender**
o Male
o Female
o Other

**What is your age?**
o (4) Under 25
o (5) Between 25 and 34
o (6) Between 35 and 44
o (7) Between 45 and 54
o (8) Between 55 and 64
o (9) Between 65 and 74
o (10) 75 and over

Page Break

In some of the following questions, we will ask you to think about the percentage chance that something will happen in the future. Your answers can be between 0 and 100, where 0 means there is no chance at all, and 100 means there is absolute certainty.

**For example, numbers like:**
2 or 5 percent mean "almost no chance"
18 percent means "not much chance"
47 or 52 percent describe a situation where there is "an equal chance that it will happen or not happen"
83 percent means "very good chance"
95 or 98 percent mean "almost certain"

Page Break



An economic recession means a decline in local economic output, such that in a certain period (such as a quarter) fewer products and services are produced in the economy compared to the previous period. An economic recession is usually reflected in a decline in average wage, employment, industrial output, and volume of products and services sold.

In your opinion, is the probability of an economic recession in the third quarter (July to September) of 2024 higher or lower than 5% 40%?[19]

o Higher than 5% 40%
o Lower than 5% 40%



In your opinion, what is the probability that there will be an economic recession in the third quarter of 2024 (July to September)?

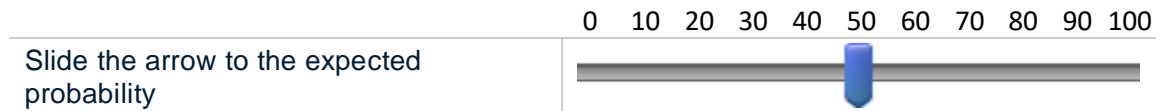



What do you think will happen to the unemployment rate in Israel in the next 12 months?
o Unemployment will increase greatly
o Unemployment will increase somewhat
o Unemployment will remain at the same level
o Unemployment will decrease somewhat
o Unemployment will decrease greatly

What do you think is the probability that in 12 months from today the unemployment rate in Israel will be higher than today?

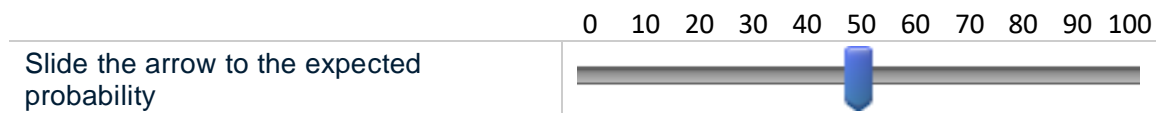

Looking ahead, do you think you (and family members living with you) will be in a better or worse financial situation 12 months from today compared to your current financial situation?
o Much worse
o Worse
o Slightly worse
o About the same
o Slightly better
o Better
o Much better

---

[19] The value in black font is the low anchor value, and the value in red is the high anchor value.



What do you think is the probability that you will lose your main job in the next 12 months?

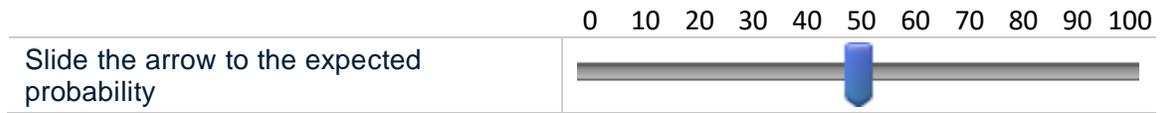

| | 0 10 20 30 40 50 60 70 80 90 100 |
|---|---|
| Slide the arrow to the expected probability | |

*Page Break*

We will now ask you a few questions about the risk of catching a number of infectious diseases, and about the effectiveness of vaccinations in reducing that risk. Please rank the following diseases from the most dangerous (top) to the least dangerous (bottom) to the best of your knowledge. You can rank them by "dragging" each disease name up or down.

Flu
Corona (COVID-19)
Measles

*Page Break*

The effectiveness of a vaccine is determined by the extent to which it reduces the risk of vaccinated people getting sick, compared to those who are not vaccinated. That is, a vaccine that is not effective at all reduces the risk of getting sick by 0%, and a perfectly effective vaccine reduces the risk by 100%.

In your opinion, does the flu vaccine reduce the risk of getting sick by less than or more than 10% 90%?
o Reduces the risk by less than 10% 90%
o Reduces the risk by more than 10% 90%

*Page Break*

To the best of your knowledge, by what percentage (%) does the flu vaccine reduce the risk of getting the flu?

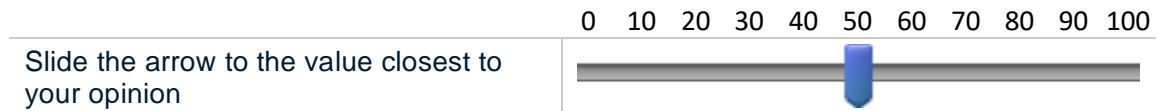

| | 0 10 20 30 40 50 60 70 80 90 100 |
|---|---|
| Slide the arrow to the value closest to your opinion | |

*Page Break*

What is the probability that you will choose to get vaccinated against the flu next year, where 0% means you will definitely not get vaccinated, and 100% means you will definitely get vaccinated?

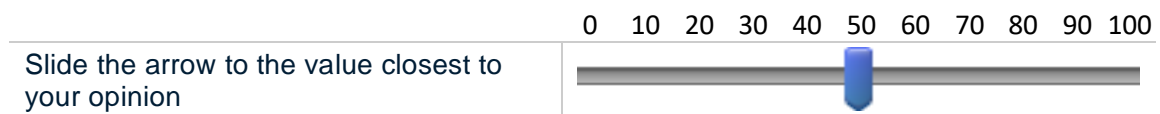

| | 0 10 20 30 40 50 60 70 80 90 100 |
|---|---|
| Slide the arrow to the value closest to your opinion | |





Were you vaccinated against the flu last winter?
o Yes
o No

Did you get the flu last winter?
o Yes
o No



Now we will ask you about activities to promote various social causes, and about the willingness of citizens to donate to such causes. Below is a list of nonprofit organizations operating in Israel. Please rank them according to the importance of their activity in your eyes from the most important (top) to the least important (bottom). You can rank by "dragging" each organization name up or down.

o Yad Sarah (lending service for medical equipment)
o Animal Welfare Association (caring for abandoned dogs and cats)
o The Society for the Protection of Nature (conservation and cultivation of nature)
o The Association for Civil Rights (protection of human rights)
o AKIM (support for people with intellectual disabilities)



Do you think that the average Israeli household donates more or less than NIS 120 (NIS 12,000) per year to nonprofit organizations and social causes?
o (1) More than NIS 120 NIS 12,000
o (2) Less than NIS 120 NIS 12,000



What do you think is the average annual donation in NIS of an Israeli household to nonprofit organizations and social causes? Please specify the amount in NIS here: _________



Assuming you win a cash prize of NIS 100,000, what is the amount of the prize that you would be willing to donate to a nonprofit organization or social cause of your choice? Please specify the amount here: _________



**II. Expectations for the national economy over time.**

### Figure A1: Expectations regarding the national economy

**How do you expect the general economic situation in this country to develop over the next 12 months? (Israeli Central Bureau of Statistics)**

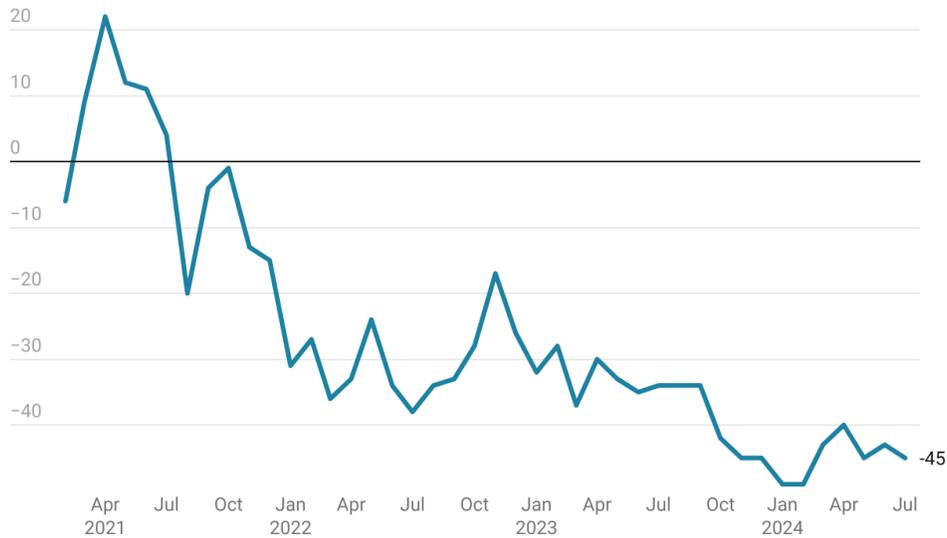

Created with Datawrapper

Note: This measure is one of the components of the Consumer Confidence Index, measured on a monthly basis by the Israel Central Bureau of Statistics. This measure is calculated by the weighted difference between the share of positive (anticipating economic improvement) and the share of negative (anticipating economic downturn) responses.



**III. Donation experiment: Multivalued anchors**

The donation experiment included 49 anchor values, from 0 to 1,000,000 INS, in an approximately logarithmic scale. The set of values was: 0, 5, 10, 15, 20, 30, 40, 50, 75, 100, 120, 150, 200, 250, 300, 400, 500, 600, 700, 800, 900, 1000, 1200, 1500, 200, 3000, 4000, 5000, 6000, 7000, 8000, 9000, 10000, 12000, 15000, 20000, 25000, 30000, 35000, 40000, 45000, 50000, 75000, 100000, 120000, 150000, 200000, 500000, 1000000. Figure A3 presents the approximate logarithmic distribution of this set.

Figure A3: The set of multivalued anchors values of the household donation experiment and their natural log transformation.

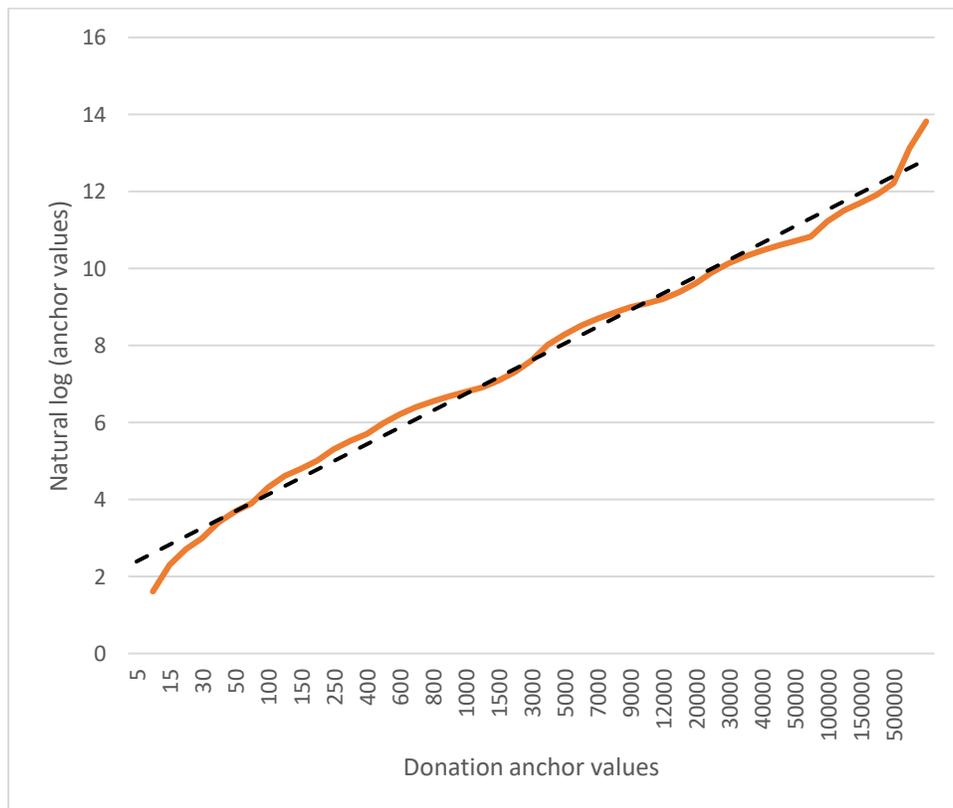



## IV. Further analysis of the decay of anchoring effects

The distribution of observations across time in the follow-up study is bimodal and ranges between 5 to 14 days (see Figure A4). To further assess the nature of the temporal decay of the anchoring effects, we estimated them separately: the first cluster of observations took place between days 4 and 9 ($M = 6.96$ days) and the second cluster of observations took place between days 10 and 14 ($M = 10.78$ days). The results show no significant difference between the relatively short- and long-delayed anchoring effects for either the recession expectation experiment or the household donation experiment.

Figure A4: Temporal distribution of lagged anchoring effects

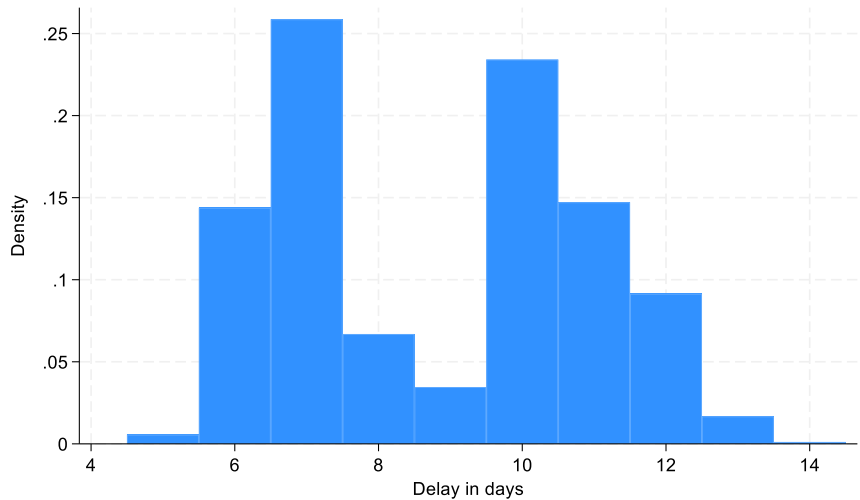

Figure A5: Decay of anchoring effects.

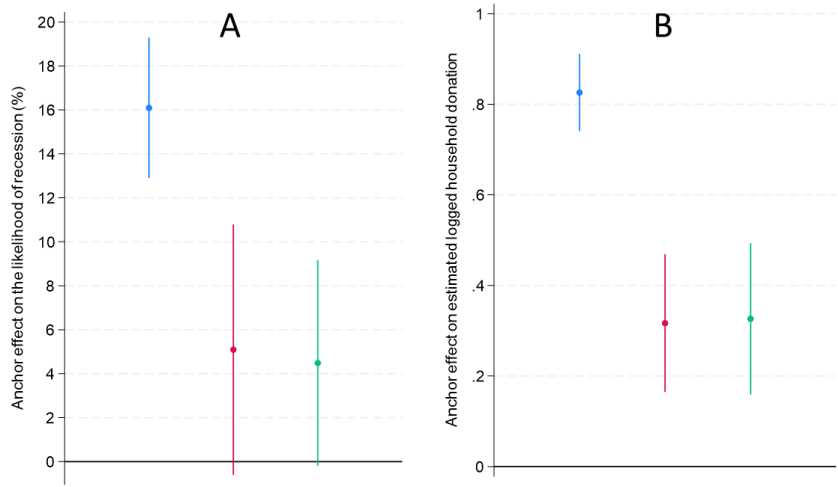

*Notes*: Panels A and B present the instantaneous (blue), 5–9 (M=6.96) days delayed (red), and 10-14 (M=10.78) days delayed (green) anchoring effects on the likelihood of recession and estimated household donation, respectively, with 95% CIs.



**V. Anchor value selection**

<u>Anchor values for recession likelihood estimates</u>

We fitted a ($3^{rd}$-order) polynomial function to the results of the multivalued anchor experiment for recession likelihood estimates, and calculated the minimum and maximum points of this function to identify the anchor values that maximize the difference between the two group means. Lastly, based on the distribution of recession likelihood estimates with no anchor exposure (in round 2), we converted the raw anchor values to percentiles.

The function that best describes the relationship between anchor values and estimates is:

$$Y = 48.932 - 0.47027x + 0.01530x^2 - 0.00009x^3$$

where its minimum and maximum are obtained at 18.5 and 91.4, respectively.

Based on the recession likelihood estimates without anchor exposure the low valued anchor is in the $12^{th}$ percentile, and the high valued anchor is in the $95^{th}$ percentile.

<u>Anchor values for estimated mean household donation</u>

Based on the results of the multivalued anchor experiment for estimated mean household we conducted the same procedure, as detailed above.

The function that best describes the relationship between anchor values and estimates is:

$$Y = 2.2179 - 0.72892x + 0.44294x^2 - 0.05196x^3$$

where its minimum and maximum are obtained at 1 and 4.7, respectively.

Based on the yearly household donation estimates without anchor exposure (excluding outliers > 100,00) the low valued anchor is in the $1^{st}$ percentile, and the high valued anchor is in the $95^{th}$ percentile.